\documentclass[10pt]{article}
\usepackage{graphicx, amsmath, amssymb, amsthm, hyperref}
\usepackage[a4paper, margin=1in]{geometry}

\title{Single-Pixel Imaging Technology in Holographic Microscopy}
\author{Arthur G. Fedorov\\ \small North-Eastern Federal University, 58 Belinsky Str., Yakutsk, 677000 Russia\\ \small Corresponding author: \texttt{ag.fedorov@s-vfu.ru}}
\date{\today}

\begin{document}
\maketitle

\begin{abstract}
We propose a holographic microscopy method based on single-pixel imaging technology (HM-SPI). We used a holographic microscopy method based on in-line Gabor holography. In single-pixel imaging technology, cyclic binary masks and amplitude-phase masks are used instead of cyclic Hadamard masks. These masks are generated using the quadratic residue and twin-prime techniques. Numerical results are presented for both cases.

Unlike the traditional approach of using DMD technology, our model considers a photodetector as a single-pixel detector. We propose a method based on the fast Fourier transform (FFT) algorithm to reconstruct the original field, which has a computational complexity of $O(NlogN)$.
This approach opens up prospects for the development of compact holographic systems capable of operating across a wide spectral range and under limited computational resources.
\end{abstract}

\section{Introduction}
Single-pixel imaging (SPI) technology is based on the registration of radiation intensity as a one-dimensional data array using photodetectors or DMD technology [1-5]. This approach relies on sequential measurement of the light intensity passing through various spatial/phase masks [6], followed by the reconstruction of an image from the one-dimensional matrix into a two-dimensional data matrix.

SPI is a promising technique in areas where the use of conventional array detectors is challenging, impractical, or expensive. For example, it is particularly useful in autonomous probes and unmanned aerial vehicles, where power consumption, size, and weight are critical factors. Another important application of this technology is its ability to operate in a wide range of wavelengths [7-10] and under low light conditions [11, 12]. The latter is especially significant in biological studies [7], where a nondestructive investigation method is crucial.

As noted above, there are two types of cyclic masks -- spatial (S-matrices) and phase (C-matrices) -- that are derived from cyclic Hadamard matrices. Cyclic Hadamard masks are constructed from sequences whose elements take values of ``+1'' and ``-1'' [13, 14]. The distinguishing feature of S-matrices is that their elements are filled with values of ``1'' and ``0'', whereas C-matrices use ``$i$'' and ``$-i$'' instead of ``+1'' and ``-1'', respectively.Therefore, both S and C matrices can be constructed based on algorithms for cyclic Hadamard matrix generation [15]. The use of phase-encoded cyclic matrices is particularly relevant in applications where the preservation of phase information is essential.

Holographic microscopy has a wide range of applications. This method has gained widespread adoption in both electron microscopy [16, 17] and optical microscopy [18] due to its relative ease of implementation. One of the key directions in digital holography is the development of compact and portable digital holographic microscopes. Additionally, these microscopes serve as the foundation for the creation of autonomous probes, such as those used to study microplastic content in aquatic environments [19, 20].

Single-pixel imaging has been successfully applied in holographic microscopy to record and reconstruct amplitude-phase information using spatially structured masks in combination with photodetectors or DMD technology. For example, visualization of biological tissues [21] is based on compressive holography using DMD technology. In that study, the authors propose extracting the phase component through heterodyning. In that work, the authors propose extracting the phase component using heterodyning. One of the key aspects, according to the authors, is that each computation takes between 1 and 1.5 hours. An alternative approach to preserve phase information is proposed in [22], where the phase shift method [23] is employed. The authors of [22] note that in their current system, the resolution of the reconstructed image is limited to 16$\times$16 pixels, and the reconstruction process takes 102 seconds, primarily due to the low operating speed of the LCOS-SLM ($\sim$10 Hz).

In light of the above, methods based on single-pixel imaging demonstrate significant potential in simplifying optical systems and reducing their cost, which is particularly important for the development of compact and portable holographic devices. However, the challenge of optimizing this technology remains relevant to ensure the highest accuracy in amplitude-phase reconstruction and to improve reconstruction speed.

The aim of this work is to propose a holographic microscopy method based on single-pixel imaging technology (HM-SPI). We present numerical modeling algorithms for the application of single-pixel imaging technology in holography. An analysis of the feasibility of complex-amplitude image reconstruction is conducted using phase masks and the phase-shifting method. The algorithm developed supports both binary cyclic masks and phase masks, ensuring versatility for various applications. The proposed method is easily adaptable to different tasks and wavelength ranges. An additional advantage is its simplicity and practicality, allowing efficient implementation even on resource-limited computational systems and enabling integration into compact, portable holographic devices.

\section{Methods}
The scheme considered in this study is shown in Fig. 1. It is assumed that a plane wave (Fig. 1-1) propagates from left to right, passing through a cyclic mask (Fig. 1-2) and interacting with the object under investigation (Fig. 1-3). The registration is performed using a single-pixel detector (e.g., a photodetector) (Fig. 1-4) located at a certain distance L from the object plane.
\begin{figure}[h]
\centering \includegraphics{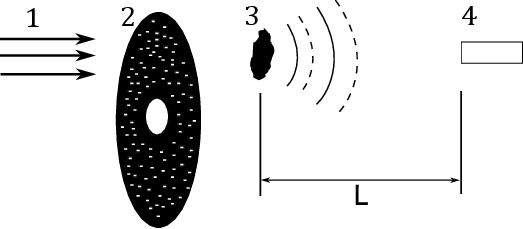} 
\caption{Wave propagation scheme: 1) Incident wave; 2) Cyclic mask; 3) Object; 4) Photodetector.}
\end{figure}

Cyclic matrices can be constructed based on Hadamard matrices [24], and there are considered to be three possible methods [13]: the quadratic residue method, the twin prime method, and the maximum-length linear feedback shift register method.  In this study, we focus on the quadratic residue and twin prime methods:
\begin{enumerate}
    \item \textit{Quadratic Residue Method}. The elements of the sequence $s$  can be defined by the following rule:
\begin{equation}
s_i =
\begin{cases} 
    1, & \text{if } i = \left(\frac{p-1}{2}\right)^2 \mod p \\
    0, & \text{otherwise}
\end{cases},
\end{equation}
where $p$ is an odd prime number, $1 \leq i \leq (p-1)$, and $s_0 = 0$. The sequence $s$ and all its cyclic shifts form an S-matrix. According to this definition, the dimension $N$ of the resulting cyclic matrix must satisfy the following condition: $p\equiv
N=4k+3$, where $k$ is an integer.
    \item \textit{Twin Prime Method}. The second approach to construct cyclic masks is based on the use of twin prime pairs p and $p+2$ (known values include $p=3, 5, 7, 11, 13, 17, 19, 29, …$) [25]. In this case, the dimension of the S-matrix is determined as $N=p(p+2)$. The sequence $s$ for $i=0,1,…,l-1$ is defined as follows:
\begin{equation}
s_i =
\begin{cases} 
    \left(\frac{i}{p}\right) \left(\frac{i}{p+2}\right), & \text{if } i \text{ is not a multiple of } p \text{ or } p+2 \\
    0, & \text{if } i \neq 0 \text{ is a multiple of } p \\
    1, & \text{if } i \text{ is a multiple of } p+2, \text{ including } i = 0
\end{cases}.
\end{equation}
Here, $\left(\frac{i}{p}\right)$ is the Legendre symbol, which equals 1 if i is a quadratic residue modulo $p$ and 0 otherwise. The same rule applies for $\left(\frac{i}{p+2}\right)$.
\end{enumerate}

Similarly, using expressions (1) and (2), C-matrices can be formed by replacing the element values ``1'' and ``0'' with ``$i$'' and ``$-i$'' respectively. As an example, let us consider the S-matrices obtained using expression (1):

\[
S_{11}^{l} =
\left[
\begin{array}{ccccccccccc}
0 & 1 & 0 & 1 & 1 & 1 & 0 & 0 & 0 & 1 & 0 \\
1 & 0 & 1 & 1 & 1 & 0 & 0 & 0 & 1 & 0 & 0 \\
0 & 1 & 1 & 1 & 0 & 0 & 0 & 1 & 0 & 0 & 1 \\
1 & 1 & 1 & 0 & 0 & 0 & 1 & 0 & 0 & 1 & 0 \\
1 & 1 & 0 & 0 & 0 & 1 & 0 & 0 & 1 & 0 & 1 \\
1 & 0 & 0 & 0 & 1 & 0 & 0 & 1 & 0 & 1 & 1 \\
0 & 0 & 0 & 1 & 0 & 0 & 1 & 0 & 1 & 1 & 1 \\
0 & 0 & 1 & 0 & 0 & 1 & 0 & 1 & 1 & 1 & 0 \\
0 & 1 & 0 & 0 & 1 & 0 & 1 & 1 & 1 & 0 & 0 \\
1 & 0 & 0 & 1 & 0 & 1 & 1 & 1 & 0 & 0 & 0 \\
0 & 0 & 1 & 0 & 1 & 1 & 1 & 0 & 0 & 0 & 1
\end{array}
\right],
S_{11}^{r} =
\left[
\begin{array}{ccccccccccc}
0 & 1 & 0 & 1 & 1 & 1 & 0 & 0 & 0 & 1 & 0 \\
0 & 0 & 1 & 0 & 1 & 1 & 1 & 0 & 0 & 0 & 1 \\
1 & 0 & 0 & 1 & 0 & 1 & 1 & 1 & 0 & 0 & 0 \\
0 & 1 & 0 & 0 & 1 & 0 & 1 & 1 & 1 & 0 & 0 \\
0 & 0 & 1 & 0 & 0 & 1 & 0 & 1 & 1 & 1 & 0 \\
0 & 0 & 0 & 1 & 0 & 0 & 1 & 0 & 1 & 1 & 1 \\
1 & 0 & 0 & 0 & 1 & 0 & 0 & 1 & 0 & 1 & 1 \\
1 & 1 & 0 & 0 & 0 & 1 & 0 & 0 & 1 & 0 & 1 \\
1 & 1 & 1 & 0 & 0 & 0 & 1 & 0 & 0 & 1 & 0 \\
0 & 1 & 1 & 1 & 0 & 0 & 0 & 1 & 0 & 0 & 1 \\
1 & 0 & 1 & 1 & 1 & 0 & 0 & 0 & 1 & 0 & 0
\end{array}
\right].
\]
As can we see, each subsequent row of $S_{11}$ is obtained by shifting the previous row by one cell. It should be noted that in practical implementation, the direction of the shift (left $S_{11}^{l}$ or right $S_{11}^{r}$) must be taken into account, as it will determine the rotation method of the cyclic mask. Graphically, $S_{11}^{l}$ and $S_{11}^{r}$ are presented in Fig. 2.
\begin{figure}[h]
\includegraphics[width=1\textwidth]{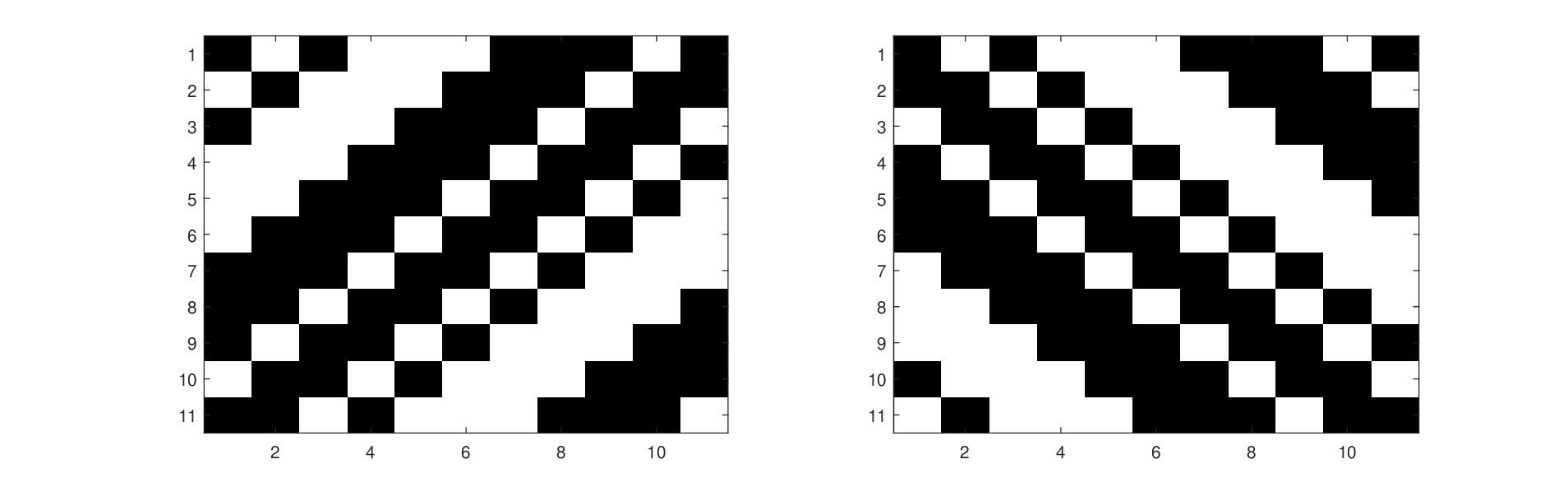} 
\caption{ Graphical representation of $S_{11}$: $S_{11}^{l}$ and $S_{11}^{r}$, respectively.}
\end{figure}

As mentioned above, circular cyclic masks in the form of disks are most commonly used in single-pixel imaging technology (Fig. 1–2) [26]. The disk rotation can be achieved using a controlled stepper motor, which must be calibrated so that each step corresponds to a separate fragment of the cyclic S-matrix. As a result, the total recorded intensity can be represented as a vector $v=(1,N)$, where $N$ corresponds to the total number of discrete angular steps of the circular disk. In other words, each step corresponds to a separate intensity measurement obtained by the photodetector, allowing the formation of a complete measurement matrix for subsequent image reconstruction.
\begin{figure}[h]
\includegraphics[width=1\textwidth]{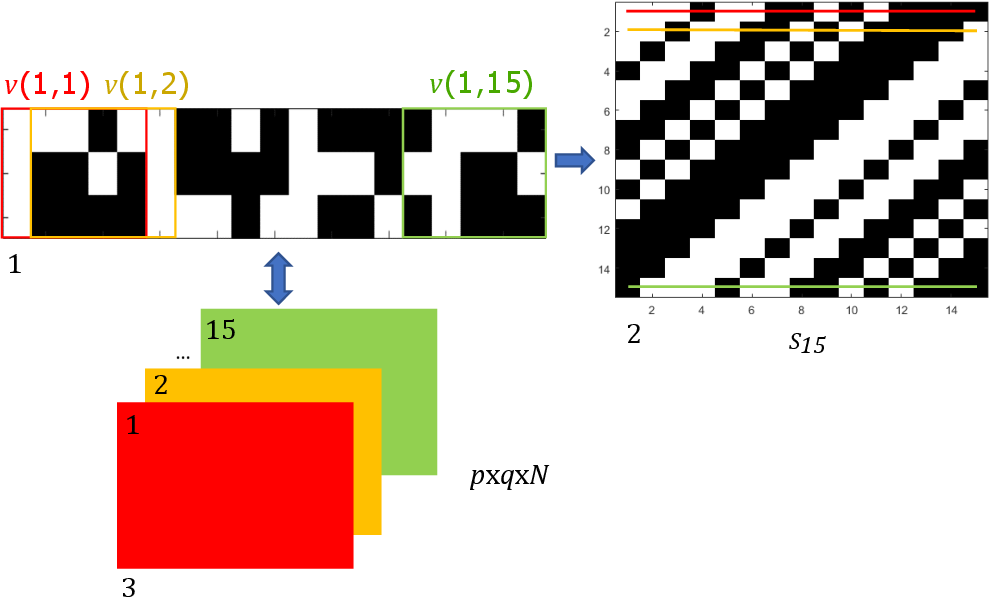} 
\caption{Schematic representation of $S_{15}$: 1) Visualization of sequence $v$; 2) $S_{15}$ obtained by twin prime method; 3) Matrix representation of sequence $v$.}
\end{figure}

In Figure 3-2, an example of a cyclic matrix $S_{15}$ is presented, obtained using the twin prime method of the form (2) with $p=3$, $q=p+2$, and $N=p\cdot q$. As a result, each value $v_i$ represents the total intensity of the submatrix $p\times q\times i$ for $i=1,N$ (Fig. 3-1). In this study, for convenience in modeling circular cyclic masks, a submatrix in the form of squares is considered (Fig. 3-1). In this formulation, the S-matrix represents a composite data matrix, where each row corresponds to a separate fragment of the intensity recorded by the photodetector.

As noted earlier, one of the methods for recording amplitude-phase information is the phase-shifting method, which involves the use of additional reference beams with different phase shifts. In this study, the inline Gabor scheme [27] (Fig. 1) is considered, where the use of a reference wave is an integral part of the holographic image formation process. Moreover, employing the same reference wave with different phases does not violate the conditions for hologram formation, making this approach the most preferable for holography applications.

The intensity value, taking into account the phase information, can be represented as:
\begin{equation}
V = \left| U U^* \right|,
\end{equation}
where the superscript ``$*$'' denotes complex conjugation, and $U$ is defined as:
\[
U=O+R,
\]
where $O$ is the object wave, and $R = A_r \exp(i\varphi)$ is the reference wave. The amplitude of the reference wave $A_r$ can be adjusted; for example, in [22], it is suggested that $A_r$ be set equal to the power of a single pixel of the object wave. In this study, a scheme with a reference wave amplitude of $A_r = \frac{1}{2}$ is considered, based on four phase shifts:
\[
\varphi_1 = 0, \quad \varphi_2 = \frac{\pi}{2}, \quad \varphi_3 = \pi, \quad \varphi_4 = \frac{3\pi}{2}.
\]
In accordance with the phase shifts, we have four reference waves:
\begin{equation}
    R_1 = A_r \exp(i\varphi_1), \quad R_2 = A_r \exp(i\varphi_2), \quad 
    R_3 = A_r \exp(i\varphi_3), \quad R_4 = A_r \exp(i\varphi_4).
\end{equation}
Taking (4) into account, we obtain four values for $V$: $V_1,V_2,V_3,V_4$. Then, according to (3), the following expression can be written:
\[
\tan(\varphi_{\text{tot}}) = \frac{V_2 - V_4}{V_1 - V_3}.
\]
As a result, the expression for the amplitude-phase information values can be written as follows:
\begin{equation}
    v = A_{\text{tot}} \exp(i\varphi_{\text{tot}}),
\end{equation}
where $A_{\text{tot}} = \sqrt{\frac{V_1 + V_2 + V_3 + V_4}{4}}$. Next, the reconstruction of the original field $x$ from the values of $v$ can be performed as follows:
\begin{equation}
    x = C^{-1} v,
\end{equation}
where the superscript ``-1'' denotes the inverse matrix, and $C$ generally represents a certain transformation matrix. To compute $C^{-1}$, numerical algorithms can be used, such as pseudo-inverse matrix methods, regularized approaches, or iterative optimization methods.

To accelerate the process of reconstructing the original field, we propose an approach based on the Fast Fourier Transform (FFT) algorithm. For this, we utilize the fact that the matrix $S$ is cyclic, where each subsequent row is obtained by a cyclic shift of the previous one. In this case, to compute $C^{-1}$, we can use the property of circulant matrices [28]:
\[
C = F D F^{-1},
\]
where $F$ is the discrete Fourier transform (DFT) matrix, $F^{-1}$ is the inverse DFT matrix and $D$ is the diagonal matrix of eigenvalues of $S$, calculated as:
\[
D = \operatorname{diag}(\mathcal{F}[c]),
\]
where $\mathcal{F}$ denotes the operation of applying the FFT algorithm and $c$ represents the sequence corresponding to the first row of the matrix $C$. Then,
\begin{equation}
    C^{-1} = F D^{-1} F^{-1}.
\end{equation}
Therefore, expression (6), taking into account (7), can be rewritten as follows:
\[
x = C^{-1} v = F D^{-1} F^{-1} v,
\]
or,
\begin{equation}
    x = \mathcal{F}^{-1} \left[ D^{-1} \mathcal{F}[v] \right],
\end{equation}
where $\mathcal{F}^{-1}$ denotes the inverse FFT. The application of the FFT algorithm in expression (8) reduces the computational complexity from $O(N^3)$, which is characteristic of the direct solution of equation (6), to $O(NlogN)$, making the proposed approach more efficient for large values of $N$.

As a result, from expression (8) we obtain the intensity value vector $x$, which can be transformed into a two-dimensional distribution $I$ corresponding to the holographic image.

\section{Results}

In this study, the scheme presented in Fig. 1 is considered. It is assumed that a plane wave propagates from left to right. As mentioned above, the cyclic mask is represented as a submatrix of dimensions $p\times q=101\times 103$, with the total number of layers being $N\equiv 101\times 103=10403$. The cyclic submatrix was modeled using the twin prime method. Then, considering that the S-matrix is square, its dimensions will be: $N\times N=10403\times 10403$.

The USAF image is considered an object and its dimension matches those of the submatrix.
\begin{figure}[h]
\includegraphics[width=1\textwidth]{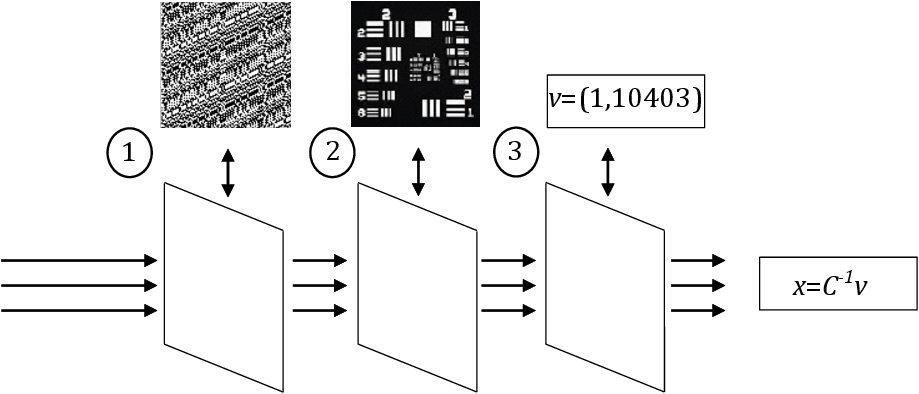} 
\caption{Wave propagation modeling.}
\end{figure}

The main stages of numerical modeling for the values $v=(1,N)$ are presented in Fig. 4. A plane wave with a wavelength of $\lambda=500\times 10^{-9}$ m passes through the cyclic mask (Fig. 4-1), then interacts with the object (Fig. 4-2), after which the recorded field is captured as a vector $v$, containing intensity values (Fig. 4-3). The numerical modeling of wave propagation was performed using the angular spectrum method [29].
\begin{figure}[h]
\includegraphics[width=1\textwidth]{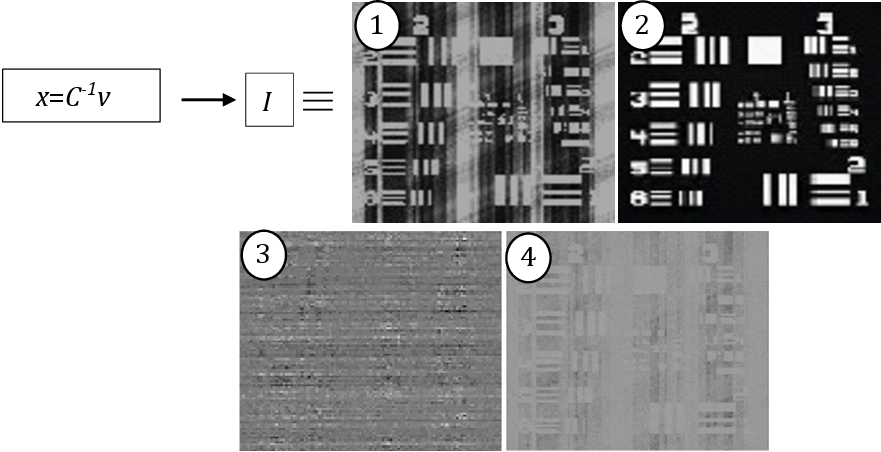} 
\caption{Reconstruction results: 1) Image obtained using our reconstruction algorithm (processed for better visualization); 2) Original image; 3) Phase information based on S-matrix; 4) Phase information based on C-matrix.}
\end{figure}

To reconstruct the original field $x$, expression (6) is applied, with its two-dimensional distribution $I$ presented in (Fig. 5-1). As seen in the figure, the distribution $I$ closely corresponds to the original image (Fig. 5-2). However, the phase information varies depending on the chosen approach.
In Figure 5-3, the phase information reconstructed using the binary cyclic S-matrix is shown, while Figure 5-4 present the phase information obtained using the C-matrix. As observed, in the first case, the phase information is almost entirely lost, making it impossible to form a holographic image. In contrast, reconstruction using the C-matrix preserves the phase information, confirming the feasibility of this method for holography applications.

\section{Conclusions}
This work demonstrates the feasibility of integrating single-pixel imaging technology into holographic microscopy. The numerical experiments conducted confirmed the effectiveness of using Hadamard cyclic masks as a basis for recording intensity, followed by the reconstruction of a complex amplitude image.

A numerical modeling algorithm has been developed that incorporates methods for generating S- and C-matrices, as well as their application for holographic image reconstruction.
The proposed reconstruction method using the Fast Fourier Transform significantly reduced the computational complexity of the problem, making the approach suitable for handling large datasets.
Thus, the study conducted  confirms that single-pixel imaging can be successfully applied in holographic microscopy for phase-amplitude information reconstruction.
The methods and algorithms proposed in this work can be utilized for the preliminary evaluation and analysis of the feasibility of developing compact and autonomous low-power holographic systems capable of operating across different spectral ranges.

\end{document}